\documentclass[aps,njp,superscriptaddress]{revtex4-2} 
\usepackage{amssymb,amsmath,amsfonts,graphicx,epsf,bm} 
\usepackage{color}
\usepackage{amsthm}

\begin{document}

\title{Dimensional measures of generalized entropy}

\author{Vladimir Zhdankin}
\email{vzhdankin@flatironinstitute.org}
\affiliation{Center for Computational Astrophysics, Flatiron Institute, 162 Fifth Avenue, New York, NY 10010, USA}

\date{\today}

\begin{abstract}
Entropy is useful in statistical problems as a measure of irreversibility, randomness, mixing, dispersion, and number of microstates. However, there remains ambiguity over the precise mathematical formulation of entropy, generalized beyond the additive definition pioneered by Boltzmann, Gibbs, and Shannon (applicable to thermodynamic equilibria). For generalized entropies to be applied rigorously to nonequilibrium statistical mechanics, we suggest that there is a need for a physically interpretable (dimensional) framework that can be connected to dynamical processes operating in phase space. In this work, we introduce dimensional measures of entropy that admit arbitrary invertible weight functions (subject to curvature and convergence requirements). These ``dimensional entropies'' have physical dimensions of phase-space volume and represent the extent of level sets of the distribution function. Dimensional entropies with power-law weight functions (related to R\'{e}nyi and Tsallis entropies) are particularly robust, as they do not require any internal dimensional parameters due to their scale invariance. We also point out the existence of composite entropy measures that can be constructed from functionals of dimensional entropies. We calculate the response of the dimensional entropies to perturbations, showing that for a structured distribution, perturbations have the largest impact on entropies weighted at a similar phase-space scale. This elucidates the link between dynamics (perturbations) and statistics (entropies). Finally, we derive corresponding generalized maximum-entropy distributions. Dimensional entropies may be useful as a diagnostic (for irreversibility) and for theoretical modeling (if the underlying irreversible processes in phase space are understood) in chaotic and complex systems, such as collisionless systems of particles with long-range interactions.
\end{abstract}


\maketitle

\section{Introduction}

Characterizing entropy is an issue of central importance for statistical physics and information theory. Here, we use the term ``entropy'' loosely to refer to a measure of the smoothness (or featurelessness) of the distribution function $f(X)\ge 0$ of a sampling variable $X \in {\mathcal P}$ at a specified (discrete) binning scale, such that smoother states have higher entropy. Here, ${\mathcal P}$ is the phase space consisting of a set of measurable properties; for example, ${\mathcal P} \subset \mathbb{R}^6$ for a population of particles with three-dimensional coordinates in both space and momentum. Entropy is often used as an indicator of irreversibility, randomness, mixing, and number of microstates. In another sense, entropy can be interpreted as a measure of the dispersion of the sampling variable, when expressed as the ``entropy power'' \citep[e.g.,][]{shannon_1948, campbell_1966, bobkov_chistyakov_2015, chen_etal_2016, jizba_etal_2016, zhdankin_2022b}.

It is known that for physical systems in thermal equilibrium, the relevant entropic quantity is the classical Boltzmann-Gibbs-Shannon (BGS) entropy,
\begin{align}
S_{\rm BGS} = - k_B \int dX \, f \log{\left(\frac{f}{f_{\rm ref}}\right)} \, , \label{eq:sbgs1}
\end{align}
where $k_B$ and $f_{\rm ref}$ are constants required for proper normalization. The BGS entropy uniquely characterizes the kinetic structure of a thermal system that lacks long-range correlations. More broadly, however, it represents only one of a countless number of possible measures for entropy. It has been challenging to develop a widely accepted formalism of entropy applicable to ``nonequilibrium'' statistical mechanics, where systems are out of local thermal equilibrium. Nonequilibrium statistical systems may display irreversibility and convergence to certain long-lasting states, but {there is no expectation for them to be} fully characterized by $S_{\rm BGS}$.

Many mathematical definitions of ``generalized'' entropies have been suggested for applications in information theory, statistical mechanics, and applied sciences \citep[e.g.,][]{renyi_1961, campbell_1966, havrda_charvat_1967, tsallis_1988, kaniadakis_2002, tsekouras_tsallis_2005, hanel_thurner_2011}; we point the reader to recent reviews \citep{amigo_etal_2018, ilic_etal_2021} and discussion of perceived limitations \citep{cho_2002, presse_etal_2013, jizba_korbel_2019}. The number of proposed generalized entropies (and their defining tenets) is overwhelming, without a clear unifying scheme. A major limitation of invoking these generalized entropies for problems in nonequilibrium statistical physics is that their connection to dynamical processes is indirect, due to the formalisms not having a straightforward physical interpretation. Thus, for a given problem, the question of ``which generalized entropy, if any, is relevant for this problem?'' typically does not have an easy or unique answer. As a consequence, the free indices or functional forms associated with generalized entropies are typically treated as unconstrained parameters. There are few exceptions where the relevant entropies can be calculated from first principles, such as the semiclassical momentum distribution of atoms in optical lattices \citep{lutz_2003, douglas_etal_2006, lutz_renzoni_2013} {or the charged particle distribution in a plasma subject to an imposed bath of electromagnetic field fluctuations \citep[e.g.,][]{hasegawa_etal_1985, ma_summers_1998, yoon_2014}}.

For applications in nonequilibrium statistical mechanics, we argue that it is {beneficial} to formulate generalized entropies in a way that connects to irreversible processes occurring in particular regions of phase space: resonances, phase mixing, cascades, dissipative structures, etc. Such a connection would enable generalized entropies to have a predictive capability based on first principles, rather than simply being a diagnostic or motivated by universality arguments. {This connection can be established once generalized entropies are expressed} in a ``dimensional'' framework. In particular, when $X$ is taken to be a quantity in $d$-dimensional phase space $\mathbb{R}^d$ with physical dimensions (e.g., length, time, or mass) in each coordinate, the distribution $f$ has dimensions of inverse phase-space volume (where ``volume'' here refers to the $d$-volume formed as a product of all components, which may have differing physical dimensions). Any operations on $f$, including the calculation of entropies, should be performed in a dimensionally consistent manner \citep{barenblatt_1987}, which does not seem to have been a priority in the literature. Dimensional consistency can be viewed as a passive symmetry that must be universally satisfied by any physical model; this is sometimes described in the literature as units covariance \citep[e.g.,][]{villar_etal_2023}.

In this work, we take a step toward making generalized entropy more accessible and useful to the physics community, by presenting a dimensional formulation of generalized entropy that may be directly related to processes operating in the phase space. We introduce ``dimensional entropies'' that are built from arbitrary invertible weight functions (subject to curvature and convergence requirements), and thus are equivalent to many previously proposed entropies as special cases. We argue that entropies derived from power-law weight functions are particularly useful due to their scale invariance; these are fundamentally equivalent to the entropies by R\'{e}nyi, Campbell, Havrda-Charv\'{a}t, and Tsallis  \citep{renyi_1961, campbell_1966, havrda_charvat_1967, tsallis_1988}, differing only in the regularization of the phase-space integral. Our formulation is dimensionally consistent without introducing {\it ad-hoc} normalization coefficients (such as $k_B$ and $f_{\rm ref}$ in Eq.~\ref{eq:sbgs1}), yielding a quantity that has the same physical dimensions as the phase-space volume. The dimensional entropies are well suited as a diagnostic for irreversibility, and may be applied to elucidate the role of generalized entropy in a range of physical systems, such as collisionless plasmas \citep[e.g.,][]{zhdankin_2022, zhdankin_2022b, ewart_etal_2022}, strongly coupled plasmas \citep{wong_etal_2018}, gravitationally interacting systems \citep[e.g.,][]{lynden-bell_1967, tremaine_etal_1986, esilva_etal_2017, esilva_etal_2019b, gruzinov_etal_2020}, collisionless systems with long-range interactions \citep{chavanis_2004, chavanis_etal_2022}, classical turbulence \citep{miller_1990, robert_1991, beck_2000}, quantum processes \citep{santos_etal_2017, brunelli_etal_2018, shukla_etal_2019, landi_paternostro_2021}, chemical reactions \citep{prigogine_1978}, and biophysics \citep{brouers_sotolongo-costa_2006}.

In Section~\ref{sec:theory}, we introduce the dimensional measures of generalized entropy ($S_h$) and derive maximality conditions. We also discuss the special case of scale-invariant weight functions ($S_q$) and comment on the construction of composite entropies. In Section~\ref{sec:perturb}, we calculate the response of the dimensional entropies to perturbations of the distribution. We use the example of an exponential distribution to show that the relative change in entropy is maximized for dimensional entropies that are weighted at a scale comparable to the perturbation in phase space, confirming the link between dynamics (perturbations) and statistics (entropies). In Section~\ref{sec:apps}, we discuss potential applications of dimensional entropies for diagnosing irreversibility and for maximum entropy modeling. Finally, in Sec.~\ref{sec:conclusions}, we summarize the results and conclude.

\section{Theoretical framework} \label{sec:theory}

\subsection{Preliminaries}

Entropy is a concept that has an intuitive physical meaning as disorder, but this does not translate to a unique mathematical definition. In this work, we use the term ``entropy'' loosely to refer to a measure of the smoothness (featurelessness) of the distribution function $f : {\mathcal P} \to \mathbb{R}_{\ge 0}$ of a variable $X \in {\mathcal P}$, such that smoother states have higher entropy. The distribution is normalized by $\int dX \, f(X) = N$, with $N$ the statistical sample size; although it is common in the literature to normalize $f$ such that $N = 1$, we retain $N$ explicitly since the scaling with $N$ is of interest to certain problems~\footnote{{While we take $N$ to be a dimensionless number, the framework described in the paper is also applicable to dimensional $N$ (e.g., for a mass distribution).}}. Our formulation is applicable to phase spaces ${\mathcal P}$ that may be either discrete or continuous; in the following, motivated by statistical physics, we focus on continuous ${\mathcal P}$ and asymptotically large sample size $N \gg 1$. In practice, continuous phase-spaces are discretized by having $f$ binned at a finite measurement scale (the ``coarse-graining'' scale).

Mixing in phase space will cause $f$ to develop fine-scale structure that permeates the accessible part of ${\mathcal P}$, which upon coarse graining leads to a distribution that is featureless {(i.e., lacking ordered structure at characteristic scales)}. In this sense, entropy is maximized by a uniform distribution when considering an unconstrained dimension (e.g., space in homogeneous systems or direction in isotropic systems). When considering a constrained dimension (such as energy of particles), however, entropy may not be maximized by a unique distribution; rather, there is a broad class of possible distributions that can be considered as ``featureless'' or ``maximally smooth'', with the one that is realized in a particular statistical system depending on factors such as hidden constraints, dynamics, evolution history, etc. Regardless, maximum entropy distributions form only a small subset of all possible distributions. Our objective is then to determine a representation of entropy that will be broadly applicable while also having a physically interpretable meaning. 

Mathematically, generalized entropies are typically assumed~\footnote{This form is not completely general. For example, functionals involving multiple phase-space integrals are suggested by Ref.~\citep{esteban_morales_1995}; we return to this issue in Sec.~\ref{sec:composite}. More broadly, entropy estimators can be constructed from correlations between individual samples in the phase space \citep[e.g.,][]{kozachenko_leonenko_1987, berrett_etal_2019}, which is outside of the scope of the present work.} to be a functional of the form \citep{salicru_etal_1993}
\begin{align}
S_{H,h}[f] \equiv H{\left( \frac{1}{N} \int dX \, f h(f) \right)} \, , \label{eq:sgen}
\end{align} 
where $h:\mathbb{R}_{\ge 0} \to \mathbb{R}$ and $H:\mathbb{R} \to \mathbb{R}$ are functions that must satisfy certain properties (described below), and the integral is over the entire phase space ${\mathcal P}$. We will call $h(f)$ the weight function and $H(z)$ the regularization function. Note that Eq.~\ref{eq:sgen} defines a ``kernel-form'' entropy, while ``trace-form'' entropies assume $H(z) = z$ \citep{ilic_etal_2021}.

In previous works, it was suggested that to be a valid measure of entropy, $S_{H,h}$ must satisfy the first three Shannon-Khinchin axioms \citep{shannon_1948, khinchin_1957} (see also, e.g., \citep{amigo_etal_2018}):
\begin{itemize}
\item {\it Continuity}: $S_{H,h}[f]$ is a continuous functional of $f$.
\item {\it Maximality}: $S_{H,h}[f]$ is maximized by a uniform distribution, in the absence of any constraints.
\item {\it Expansibility}: $S_{H,h}[f]$ is unchanged if ${\mathcal P}$ is expanded by regions where $f = 0$.
\end{itemize}
The fourth Shannon-Khinchin axiom {is} strong additivity (or separability), {which states that the entropy of the joint distribution of two random variables $X,Y$ is equal to the entropy of the distribution of $X$ plus the conditional entropy of the distribution of $Y$ given $X$. In the case that $X$ and $Y$ are independent, the entropies are thus additive.} The only trace-form entropy to satisfy strong additivity is the BGS entropy. The fourth axiom will not be discussed further here. Note that alternative systems of axioms also exist, such as by Shore and Johnson \citep{shore_johnson_1980}, with interconnections discussed in Ref.~\citep{jizba_korbel_2020}. 

In contrast to most previous works, here we seek a measure of entropy that is dimensionally consistent \citep{barenblatt_1987}, an issue which is often bypassed in the literature by normalizing the quantities to a particular scale (e.g., binning scale). For example, the BGS entropy $S_{\rm BGS}$ from Eq.~\eqref{eq:sbgs1} has $H(z) = z$ and $h(f) = - N k_B \log{(f/f_{\rm ref})}$, where $f_{\rm ref}$ and $k_B$ are dimensional constants that must be prescribed. As a consequence, both the normalization and zero point of $S_{\rm BGS}$ are essentially arbitrary {(e.g., dependent on the coarse graining of the system)}. We will show that such constants are unnecessary if the entropy is formulated and interpreted in a dimensional manner.

In the following, we will assume that ${\mathcal P}$ is a dimensional phase space, which can be treated as a subset of $\mathbb{R}^d$, where $d$ is the number of {phase space} dimensions. Each coordinate $X_i$ (with $i \in \{ 1, \dots, d \}$) has a physical dimension that we denote $\dim{(X_i)}$. For example, $\dim{(X_i)}$ may {be} length, time, or mass; {it may also involve products or ratios of these (e.g., momentum has units of mass times length divided by time)}. {Applying the methods of dimensional analysis \citep{barenblatt_1987},} the physical dimensions of the $d$-volume in phase space are then a product of those of the coordinates, $\prod_{i=1}^d \dim{(X_i)}$. Since $\int dX \, f = N$ forms a dimensionless number, the distribution function has physical dimensions of inverse phase-space volume,
\begin{align}
\dim{(f)} = \prod_{i=1}^d \frac{1}{\dim{(X_i)}} \, .
\end{align}
Any dimensional measure of entropy that is constructed from $f$ must have dimensions that are consistent with $f$, by containing only the combination $\prod_{i=1}^d \dim{(X_i)}$. 

The integral $\int dX \, f h(f)$ in Eq.~\ref{eq:sgen} has physical dimensions that are determined by $h$. The dimensionality of $h$ is a degree of freedom in the definition, which will be further discussed in Sec.~\ref{sec:affine}. For invertible $h$, the inverse weight function $h^{-1}$ (not to be confused with $1/h$) provides a natural map from the dimensional space of $h$ back to a quantity with dimensions of phase space. In the next section, we will use this inverse map to construct a measure of entropy that has physical dimensions of phase-space volume, $\prod_{i=1}^d \dim{(X_i)}$.

\subsection{Dimensional entropies}

Consider any continuous, invertible weight function $h(f)$ that satisfies $f h(f)|_{f=0} = 0$ and is convex (concave) in $f h(f)$ if $h$ is an increasing (decreasing) function of $f$. We then propose to choose the regularization function $H(z) = N/h^{-1}(z)$ in Eq.~\ref{eq:sgen}. This leads to class of functionals that we will call the {\it dimensional entropies}
~\footnote{It may be conceptually simpler to consider the inverse entropy $N/S_h$ (rather than $S_h$) as the fundamental quantity, which has the same dimensions as $f$. Under this guise, maximizing $S_h$ is equivalent to minimizing the phase-space density measured by $N/S_h$.}:
\begin{align}
S_h[f]  \equiv \frac{N}{h^{-1}{\left( \frac{1}{N} \int dX \, f h(f) \right)}} \, . \label{eq:sdef}
\end{align}
The denominator in Eq.~\ref{eq:sdef} resides on the same space as the distribution $f$ (thus having the same dimensions) and resembles a weighted quasi-arithmetic mean of $f$ (or Kolmogorov expected value \citep{decarvalho_2016}), having the interpretation of being the value of $f$ at which the $f$-weighted mean of $h$ (across ${\mathcal P}$) is located. In essence, $S_h$ describes a characteristic phase-space volume occupied by $f$ in the weighted regions. Intuitively, the increase of entropy is manifest by {the spreading of the distribution $f$ across the accessible phase space, which corresponds to an} increase of the phase-space volume occupied by level sets of $f$.

Note that $N/V^*_{\mathcal P} \le h^{-1}{\left(  \int dX \, f h(f) / N \right)} \le f_{\rm max}$ (which follows from the maximality proof below), where $V^*_{\mathcal P}$ is the volume of the phase space ${\mathcal P}$ which is occupied by nonzero $f$ (which can be infinite) and $f_{\rm max} = \max{(f)}$ is the global maximum of $f$ across ${\mathcal P}$. The range of the dimensional entropies is therefore
\begin{align}
\frac{N}{f_{\rm max}} \le S_h \le V^*_{\mathcal P} \, .
\end{align} 

For the special case of a uniform distribution on a bounded domain, $f(X) = f_c \equiv N/V_{\mathcal P}$ (where $V_{\mathcal P} = \int dX$ is the total phase-space volume of ${\mathcal P}$), the dimensional entropies are degenerate, with $S_h[f_c] = N/f_c = V_{\mathcal P}$ for all choices of $h$. Thus, for the uniform distribution, $S_h$ all reduce to the total phase-space volume occupied.

{The dimensional entropies can be related explicitly to the phase-space volume contained by level sets. To do this, consider the special case of weight $h(f) = \Theta{(f-f_s)}/f$, where $f_s$ is a constant that represents the value of a level set of $f$ and $\Theta(z)$ is the Heaviside step function ($\Theta(z) = 0$ for $z<0$ and $\Theta(z) = 1$ for $z \ge 0$). In this case, the dimensional entropy reduces to the phase-space volume occupied by points with $f \ge f_s$: $S_h = \int dX \Theta{(f - f_s)}$.}

One can immediately verify that $S_h$ satisfies the Shannon-Khinchin axioms of continuity and expansibility. We now demonstrate below that it also satisfies the axiom of maximality, as long as $hf$ is convex if $h'>0$, or $hf$ is concave if $h'<0$ (here and elsewhere, primes indicate derivatives). {As normalizing a uniform random distribution requires the phase space ${\mathcal P}$ to be finite, the following proof applies to bounded phase spaces.} \\

{\bf Theorem 1 (Maximality):} Let $f(X) = f_c + \delta f(X)$, where $f_c$ is the uniform distribution on {a bounded domain} ${\mathcal P}$ and $\delta f$ is the deviation from the uniform distribution, such that $\int dX \, \delta f = 0$. Then $S_h[f] \le S_h[f_c]$ if $h$ is chosen such that $f h(f)$ is convex if $h'(f)>0$ or concave if $h'(f)<0$.

\begin{proof}[Proof of Theorem 1]
First consider convex $f h(f)$, and $h(f)$ an increasing function of $f$. Then $f h(f) \ge f_c h(f_c) + [f_c h(f_c)]' \delta f$, implying $\int dX \, f h(f)/N \ge h(f_c)$. Since $h^{-1}(z)$ is also an increasing function of $z$, we have that $h^{-1}\left(\int dX \, f h(f)/N\right) \ge h^{-1}\left(h(f_c)\right) = f_c$. Then
\begin{align}
S_h[f] = \frac{N}{h^{-1}\left(\frac{1}{N} \int dX \, f h(f)\right)} \le \frac{N}{f_c} = S_h[f_c] \, .
\end{align}

Now consider the opposite case, with $f h(f)$ concave and $h(f)$ a decreasing function of $f$. Reversing the initial inequality for the above case, $f h(f) \le f_c h(f_c) + [f_c h(f_c)]' \delta f$, so $\int dX \, f h(f)/N \le h(f_c)$. Now $h^{-1}(z)$ is a decreasing function of $z$, so again $h^{-1}\left(\int dX \, f h(f)/N\right) \ge f_c$, implying $S_h[f] \le S_h[f_c]$.
\end{proof}

\subsection{Invariance of dimensional entropies under affine transformation of weight function} \label{sec:affine}

One property of $S_h$ is that it is invariant under affine transformations of $h$ (shifts and rescalings of $h$ by constant factors), which implies that $S_h$ is independent of the physical dimensions of $h$. {This property follows from $S_h$ having the form of an inverse weighted quasi-arithmetic mean. We now provide a proof of this statement:} \\

{\bf Theorem 2:} Let $\tilde{h}(f) = A h(f) + B$ be a new weight function which is rescaled and shifted by constant values $A$ and $B$, respectively. Then $S_{\tilde{h}} = S_h$.

\begin{proof}[Proof of Theorem 2]
By definition, $\tilde{h}(f_1) = A h(f_1) + B$ for any function $f_1$, so that $f_1 = \tilde{h}^{-1}(A h(f_1) + B)$. Choosing $f_1 = N/S_h[f]$ in this identity,
\begin{align}
\frac{N}{S_h} &= \tilde{h}^{-1}\left(A h\left(\frac{N}{S_h}\right) + B\right) \nonumber \\
&= \tilde{h}^{-1}\left( A \frac{1}{N} \int dX \, f h(f) + B \right) \nonumber \\
&= \tilde{h}^{-1}\left( \frac{1}{N} \int dX \, f \tilde{h}(f) \right)  = \frac{N}{S_{\tilde{h}}} \, .
\end{align}
\end{proof}

We note that while $S_h$ is invariant with respect to rescalings of $h$, it is not (in general) invariant with respect to rescalings of $f$ (or equivalently, $N$). Only in special cases, such as the scale-invariant entropies described in Sec.~\ref{sec:powers}, does it become invariant with respect to the normalization of $f$.

\subsection{Scale-invariant dimensional entropies} \label{sec:powers}

In general, the weight function $h(f)$ may contain characteristic scales, which requires choosing dimensional parameters (having units of inverse phase-space volume) that normalize $f$. For example, for an exponential weight function, $h = \exp{(f/f_e)}$, the constant $f_e$ must be chosen with the same dimensions as $f$. The resulting value of $S_h$ will depend on such dimensional parameters, as the parameters will affect the relative weight of different regions of phase space. For an ``agnostic'' measure of entropy that does not have any internal dimensional parameters, as characteristic of a truly ``featureless'' distribution, the weight function must be scale invariant. Scale-invariant weight functions take the form of a power law, $h = f^{q-1}$, where the weight index $q > 0$ {(with $q \neq 1$)} is defined analogously to that in the R\'{e}nyi \citep{renyi_1961} and Tsallis \citep{tsallis_1988} entropies. The resulting {\it scale-invariant dimensional entropies} take the form
\begin{align}
S_q[f] \equiv S_{f^{q-1}}[f] = N \left( \frac{1}{N} \int dX \, f^q \right)^{-1/(q-1)} \, , \label{eq:sqdef}
\end{align}
where we used shorthand notation such that the subscript $q$ on $S$ implies usage of the power-law weight function. $S_q$ are essentially identical to the exponential of the R\'{e}nyi entropy, previously discussed as ``exponential entropies'' \citep{campbell_1966} or ``R\'{e}nyi entropy powers'' \citep{bobkov_chistyakov_2015, jizba_etal_2016}.

The scale-invariant dimensional entropies have the property that the rescaled distribution $\tilde{f} = C f$ with normalization $\tilde{N} = C N$ will have the same entropy as the original distribution, $S_q[\tilde{f}] = S_q[f]$. Here, $C$ is an arbitrary constant multiplier. In this sense, $S_q$ is invariant with respect to the normalization of $f$. {In contrast, rescaling the coordinate system $X \to \tilde{X}$ while keeping $N$ fixed introduces a factor of the Jacobian determinant with the distribution, so that $dX \to |\det{(J)}| d\tilde{X} $ and $f(X) \to f(\tilde{X})/|\det{(J)}|$, where $J_{ij} = dX_i/d\tilde{X}_j$ is the Jacobian matrix ($i,j = 1, ..., d$). If $|\det{(J)}|$ is independent of $\tilde{X}$ (as with a change of units), then $S_q \to |\det{(J)}| S_q$ under the rescaling of the coordinates. Therefore, $S_q$ is covariant with the phase space volume.}

Note that $S_q$ is, up to a factor of $N$, the inverse of the weighted power mean,
\begin{align}
\langle f \rangle_{w;\zeta} &\equiv \left( \frac{\int dX \, w(X) f^{\zeta}(X)}{\int dX \, w(X)} \right)^{1/\zeta} \, ,
\end{align}
with $\zeta = q-1$ and the special choice of weight $w(X) = f(X)$. As a consequence of the weighted power mean inequality, for any two ordered indices $q_1$ and $q_2$ {with finite corresponding dimensional entropies},
\begin{align}
q_1 < q_2 \implies S_{q_1} \ge S_{q_2} \, .
\end{align}
Thus, $S_q$ form an ordered continuum of values ranging from $S_{q \to 0} = V^*_{\mathcal P}$ to $S_{q \to \infty} = N/f_{\rm max}$, where $V^*_{\mathcal{P}}$ is the phase-space volume occupied with a nonzero value of $f$. 

The parameter $q$ represents a continuous degree of freedom in the entropy. There are several notable values of $q$:
\begin{itemize}
\item The limit $q \to 0$ represents the ``lowest'' weight. In this case, $S_{q \to 0} \to V^*_{\mathcal{P}}$ for all distributions $f$. This is equivalent to the max-entropy or Hartley entropy \citep{hartley_1928}.
\item At $q=1$, $S_q$ is undefined. However, the limit $q \to 1$ (from either direction) exists and can be calculated by expanding in $\epsilon = q-1$ with $|\epsilon| \ll 1$, {which evaluates to}
\begin{align}
S_{q \to 1} &= \lim_{\epsilon \to 0} N \left( \frac{1}{N} \int dX \, f^{1+\epsilon} \right)^{-1/\epsilon} \nonumber \\
&= {\lim_{\epsilon \to 0} N \left[ \frac{1}{N} \int dX \, f \exp{\left(\epsilon \log{f}\right)} \right]^{-1/\epsilon}} \nonumber \\
&= {\lim_{\epsilon \to 0} N \left[ \frac{1}{N} \int dX \, f \left( 1 +\epsilon \log{f} \right) \right]^{-1/\epsilon}} \nonumber \\
&= {\lim_{\epsilon \to 0} N \left( 1 + \epsilon \frac{1}{N} \int dX \, f  \log{f} \right)^{-1/\epsilon}} \nonumber \\
&= N \exp{\left( - \frac{1}{N} \int dX \, f \log{f} \right)} \, , \label{eq:sqto1intermediate}
\end{align}
{where we used the identity $\exp{x} = (1 + \epsilon x)^{1/\epsilon}$ in the limit $\epsilon \to 0$. Eq.~\ref{eq:sqto1intermediate} shows that $S_{q \to 1} = S_{\log{f}}$, which is equivalent to the BGS entropy power \citep{shannon_1948}.  By introducing the arbitrary constant $f_{\rm ref}$ (from Eq.~\ref{eq:sbgs1}) with the same dimensions of $f$, Eq.~\ref{eq:sqto1intermediate} can be expressed in terms of the normalized BGS entropy as}
\begin{align}
S_{q \to 1} &= \frac{N}{f_{\rm ref}} \exp{\left( - \frac{1}{N} \int dX \, f \log{(f/f_{\rm ref})} \right)} \nonumber \\
&= \frac{N}{f_{\rm ref}} \exp{\left(\frac{S_{\rm BGS}}{N k_B} \right)} \, . \label{eq:sqto1}
\end{align} 
\item For $q = 2$, the weight function is its own inverse, $h(f) = h^{-1}(f) = f$. Thus, the dimensional entropy takes a particularly simple form, $S_2 = N^2 / \left( \int dX \, f^2 \right)$. The characteristic phase-space density $N/S_2 =  \int dX \, f^2/N$ in this case needs no regularization.
\item The limit $q \to \infty$ represents the ``highest'' weight. In this case, the min-entropy is recovered, $S_{q \to \infty} = N/f_{\rm max}$. This has the interpretation of the phase-space volume that would be occupied if the distribution was a binary function, $f \in \{ 0, f_{\rm max} \}$.
\end{itemize}

\subsection{Composite dimensional entropies} \label{sec:composite}

It is worth revisiting the assumption of a kernel-form entropy in Eq.~\eqref{eq:sgen}{, which restricts the measure to a single phase-space integral}. From the infinite set of dimensional entropies $\{ S_{h} \}$, one may construct more complex functionals {(involving multiple independent phase-space integrals)} that satisfy the first three Shannon-Khinchin axioms and thus constitute valid entropies. {As this process manipulates the distribution of $\{ S_h \}$, it} can be envisioned as generating ``entropies of entropies''.

In particular, the reader may recognize that linear combinations of $S_h$ for varying weight functions $h$ will obey similar properties as any individual $S_h$ {(i.e., the first three Shannon-Khinchin axioms)}. Furthermore, since $S_h$ all have dimensions of phase-space volume, it is dimensionally consistent to add them together. In this sense, the set of $\{ S_h \}$ serves as a basis for a far broader set of entropy measures {than any individual $S_h$}. For example, given a finite set of $N_h$ weight functions $h_i(f)$ (with $i = 1, \dots, N_h$), one can write a composite entropy as
\begin{align}
S_{\{ C_i; h_i \}}[f] \equiv \sum_{i=1}^{N_h} C_i S_{h_i}[f] \, ,
\end{align} 
where $C_i > 0$ are dimensionless composition weights (with $\sum_{i=1}^{N_h} C_i = 1$ for proper normalization). {In principle, this expression can be generalized} to {an infinite} continuous subset of $\{ S_h \}$, by replacing the sum with a functional integral. {This leads to (an infinite number of) quantities that we call the {\it composite entropies}:} 
\begin{align}
S_{\{ {\mathcal G}; h \}}[f] \equiv \int Dh \, {\mathcal G}[h] S_{h}[f] \, , \label{eq:pathintegral}
\end{align}
where $Dh$ denotes integration over all functions $h$ and ${\mathcal G}[h]$ is a composition weight functional, with ${\mathcal G}[h] \ge 0$ to ensure maximality and $\int Dh \, {\mathcal G}[h] = 1$ for proper normalization. {We refer the reader to Ref.~\cite{albeverio_etal_1976} for mathematical detail on functional integrals. The choice of measure and weight functional for Eq.~\ref{eq:pathintegral} is a nontrivial degree of freedom; Eq.~\ref{eq:pathintegral} is thus mainly of conceptual interest here. In practice, it maybe be unnecessary to perform the functional integral in Eq.~\ref{eq:pathintegral} over the entire functional space of $h$, and instead restrict the integral to a subclass of functions.} {As a particular} example, for the set of scale-invariant weights $h = f^{q-1}$ with $q > 0$, we can construct the {\it composite scale-invariant entropy}:
\begin{align}
S_{\{ G; q \}}[f] \equiv N \int_0^\infty dq \, G(q) \left( \frac{1}{N} \int dX \, f^q \right)^{-1/(q-1)} \, , \label{eq:mostgen}
\end{align}
where $G(q)$ is the composition weight function (satisfying $G(q) \ge 0$ to ensure maximality), which resembles the spectral distribution of $q$ values described in Ref.~\citep{tsekouras_tsallis_2005}. Imposing the normalization condition $\int_0^\infty dq \, G(q) = 1$ will preserve the interpretation of dimensional entropy as a phase-space occupation volume. The dimensionless function $G(q)$ is otherwise arbitrary. Finally, we note that if one interprets the scale-invariant entropies $S_q$ as being connected to fractal structures, then $S_{\{ G; q \}}$ is a natural generalization to multifractal structures \citep[e.g.,][]{hentschel_procaccia_1983, alemany_zanette_1994, jizba_arimitsu_2004}.

{While the composite entropies of Eq.~\ref{eq:pathintegral} and Eq.~\ref{eq:mostgen} have a definite mathematical meaning, their relevance to physical problems is unclear}. {We} postulate that nonequilibrium statistical systems {may} evolve to maximize $S_{\{ G; q\}}$ for a function $G(q)$ that is determined by the internal (microscopic) processes, subject to appropriate constraints {(such as energy)}. For example, for a system of particles interacting through long-range forces, thermal dissipation would correspond to increasing $S_{\{ G; q \}}$ with {$G(q) = \lim_{\epsilon\to 0} \delta(q - 1 + \epsilon)$}, while high-energy dissipative processes would have $G$ peaked at $q \lesssim 1$ and low-energy dissipative processes would have $G$ peaked at $q \gtrsim 1$. Detailed modeling is required to determine {the shape of} $G(q)$ for a given system. We leave further investigation of composite entropies to future work.

\section{Perturbative response of dimensional entropies} \label{sec:perturb}

\subsection{Effect of diffusion}

In this section, we discuss the response of the dimensional entropies $S_h$ to perturbations of the distribution (which may increase or decrease entropy), as a means to connect the dimensional entropy to physical processes that deform the distribution in phase space. In particular, we imagine the distribution being perturbed out of equilibrium by the dynamics, and then relaxing to a new state. The generalized entropy that will be influenced the most by the relaxation process will depend on the properties of the perturbation and the state of the equilibrium. We seek to characterize this relationship quantitatively by performing a perturbative expansion of $S_h$.

Before doing so, we first comment on the effect of diffusion, which will smooth the distribution and thus increase all entropies. As an example, consider the diffusion equation for a continuous distribution function $f(X;t)$ in the phase space ${\mathcal P} = \mathbb{R}^d$, over times $t \in \mathbb{R}_{\ge 0}$:
\begin{align}
\frac{\partial f}{\partial t} = \frac{\partial}{\partial X_i} \left( D_{ij} \frac{\partial f}{\partial X_j} \right) \, , \label{eq:diffeq}
\end{align}
where $D(X) \in \mathbb{R}^{d\times d}$ is the (symmetric, positive-definite) diffusion matrix, and repeated indices are summed. Then
\begin{align}
\frac{d S_h}{dt} &= \frac{d}{dt} \frac{N}{h^{-1}\left(\int dX \, f h/N \right)} \nonumber \\
&=  - \frac{{h^{-1}}' \left(\int dX \, f h/N \right)}{\left[h^{-1}\left(\int dX \, f h/N \right)\right]^2} \int dX \, (f h)' \frac{\partial f}{\partial t} \nonumber \\
&= - \frac{S_h^2[f]}{N^2 h'(N/S_h[f])} \int dX \, (f h)' \frac{\partial}{\partial X_i} \left(D_{ij} \frac{\partial f}{\partial X_j} \right) \nonumber \\
&= \frac{S_h^2[f]}{N^2 h'(N/S_h[f])} \int dX \, (f h)'' \frac{\partial f}{\partial X_i} D_{ij} \frac{\partial f}{\partial X_j} \ge 0 \, , \label{eq:diff}
\end{align}
where we assumed that the integral of $(f h)' D_{ij} \partial f/\partial X_j$ on the boundary of the phase space vanishes (valid for periodic boundary conditions or if $f$ declines sufficiently quickly). We also used the inverse function theorem ${h^{-1}}'(z) = 1/h'(h^{-1}(z))$. Since $f h$ is convex (concave) if $h$ is an increasing (decreasing) function, and $D$ is symmetric positive-definite, the final expression in Eq.~\ref{eq:diff} is positive, $d S_h/dt \ge 0$. Thus, diffusion always increases the dimensional entropies, consistent with maximality.

For scale-invariant weights, $h = f^{q-1}$, Eq.~\ref{eq:diff} simplifies to
\begin{align}
\frac{dS_q}{dt} &= \frac{q S_q^q[f]}{N^q} \int dX \, f^{q-2} \frac{\partial f}{\partial X_i} D_{ij} \frac{\partial f}{\partial X_j}  \, . 
\end{align}
The growth rates of the dimensional entropies have a nontrivial dependence on the diffusion matrix $D(X)$ and the instantaneous distribution $f$. In general, it is not the BGS entropy ($q \to 1$) that is maximized.

\subsection{Perturbative expansion of dimensional entropies}

To connect the dimensional entropies to fluctuations (and thus dynamics), we now consider the effect that perturbing a distribution has on $S_h$. Physically, such a perturbation may describe energy injection into a system of particles through an interaction with a field (e.g., a resonance) or body force. Thus, we consider
\begin{align}
f(X) = f_0(X) + \delta f(X) \, ,
\end{align}
where $f_0$ is a ``background'' distribution and $\delta f$ is a perturbation that satisfies $\int dX \, \delta f = 0$ and $|\delta f| \ll f_0$ for all $X$.

Since $h$ is assumed to be smooth and continuous, we can Taylor expand (to second order in $|\delta f|/f_0$) the functions $hf$ and $h^{-1}$ to get
\begin{align}
f h(f) &= f_0 h(f_0) + [f_0 h(f_0)]' \delta f + \frac{1}{2} [f_0 h(f_0)]'' (\delta f)^2 + \cdots \, , \label{eq:taylor1} \\
h^{-1}(z) &= h^{-1}(z_0) + [h^{-1}(z_0)]' \delta z {+ \frac{1}{2} [h^{-1}(z_0)]'' (\delta z)^2}  + \cdots \nonumber \\
&= h^{-1}(z_0) + \frac{\delta z}{h'(h^{-1}(z_0))} {- \frac{1}{2} \frac{h''(h^{-1}(z_0))}{[h'(h^{-1}(z_0))]^3} (\delta z)^2} + \cdots \, , \label{eq:taylor2}
\end{align}
where we used the inverse function theorem to rewrite {$[{h^{-1}}(z_0)]' = 1/h'(h^{-1}(z_0))$}, and $z = z_0 + \delta z$ is the expanded argument of $h^{-1}$, with $|\delta z|/|z_0| \sim {\mathcal O}(|\delta f|/f_0) \ll 1$. {Note that although we expanded $h^{-1}(z)$ to second order, the first-order expansion is sufficient for the calculations below.}

Either the first-order or second-order term in Eq.~\eqref{eq:taylor1} may dominate in the calculation of $S_h[f]$, depending on the width and amplitude of $\delta f$ relative to $f_0$ (and their degree of correlation). In particular, since $\int dX \, \delta f = 0$ while $\int dX \, (\delta f)^2 > 0$, we may expect $|\int dX \, [f_0 h(f_0)]' \delta f| \ll |\int dX \, [f_0 h(f_0)]'' (\delta f)^2/2|$ if $\delta f$ is sufficiently localized such that $[f_0 h(f_0)]'$ can be approximated as constant over that region of phase space.

If the first-order term dominates, $|\int dX \, [f_0 h(f_0)]' \delta f| \gg |\int dX \, [f_0 h(f_0)]'' (\delta f)^2/2|$, then Eqs.~\ref{eq:taylor1}-\ref{eq:taylor2} can be applied to find the relative change in entropy caused by the perturbation (to first order in $\delta f/f_0$),
\begin{align}
\frac{\Delta S_h}{S_h} \equiv \frac{S_h[f] - S_h[f_0]}{S_h[f_0]} \sim - \frac{S_h[f_0]}{N^2} \frac{\int dX \, [h(f_0) f_0]' \delta f}{h'(N/S_h[f_0])} \, . \label{eq:sperturb1}
\end{align}
The right hand side of Eq.~\eqref{eq:sperturb1} can be either positive or negative, depending on how the fluctuation correlates with the weight. Intuitively, the perturbation may increase or decrease the entropy if it broadens or narrows the {peaks of the} distribution, respectively {(i.e., whether it smooths or sharpens the distribution)}. In some situations, a given $\delta f$ may lead to $\Delta S_h/S_h > 0$ for certain choices of $h$ and $\Delta S_h/S_h < 0$ for other choices of $h$. {As such a $\delta f$ causes a shift in the weight of the dominant entropies, this can be called a ``transmutation'' of entropy.} {For example, if $f_0$ is a multi-modal distribution and $\delta f$ acts to broaden some peaks and narrow other peaks, then the sign of $\Delta S_h/S_h$ will be determined by the peaks that the chosen weight function $h$ is most sensitive to.} 

If the second-order term dominates, $|\int dX \, [f_0 h(f_0)]' \delta f| \ll |\int dX \, [f_0 h(f_0)]'' (\delta f)^2/2|$, then relative change in entropy (to second order in $\delta f/f_0$) is instead
\begin{align}
\frac{\Delta S_h}{S_h}  \sim - \frac{1}{2} \frac{S_h[f_0]}{N^2} \frac{\int dX \, [h(f_0) f_0]'' (\delta f)^2}{h'(N/S_h[f_0])} \, . \label{eq:sperturb2}
\end{align}
The right-hand side is negative because $fh$ is convex (concave) for $h$ increasing (decreasing), so that the perturbation $\delta f$ decreases entropy. Intuitively, a sufficiently narrow perturbation will always decrease entropy over the smoother background distribution, as a consequence of maximality.

For power-law weights, $h = f^{q-1}$, the first-order term [Eq.~\eqref{eq:sperturb1}] becomes
\begin{align}
\frac{\Delta S_q}{S_q} = \frac{S_q[f] - S_q[f_0]}{S_q[f_0]} &\sim - \frac{q}{q-1} \frac{\int dX \, f_0^{q-1} \delta f}{N \left(N/S_q[f_0]\right)^{q-1}}  \, , \label{eq:plperturb1}
\end{align}
while the second-order term [Eq.~\eqref{eq:sperturb2}] becomes
\begin{align}
\frac{\Delta S_q}{S_q} &\sim - \frac{q}{2} \frac{\int dX \, f_0^{q-2} \delta f^2}{N \left(N/S_q[f_0]\right)^{q-1}} \, . \label{eq:plperturb2}
\end{align}
Thus, $\Delta S_q$ is a nontrivial convolution between powers of $f_0$ and $\delta f$. Note that first-order integral in Eq.~\eqref{eq:plperturb1} is such that $\Delta S_q/S_q > 0$ if $\delta f$ smooths the distribution by being positive in phase-space regions of lower density and negative in regions of higher density. Conversely, $\Delta S_q/S_q < 0$ if $\delta f$ makes the distribution more strongly peaked by being negative in regions of lower density and positive in regions of higher density.

Eqs.~\ref{eq:plperturb1}-\ref{eq:plperturb2} imply that $\Delta S_q/S_q \to 0$ in the limit $q \to 0$, which is a consequence of the perturbations not changing the occupied phase-space volume $V_{\mathcal P}^*$ by the assumption of $|\delta f|/f_0 \ll 1$ (thus $\delta f = 0$ in empty regions of ${\mathcal P}$). On the other hand, in the limit $q \to \infty$, $\Delta S_q/S_q$ captures changes in the peak of $f$. Whether the extremum of $\Delta S_q/S_q$ is at $q \to \infty$ or at finite $q$ depends on the scaling and correlations of $f_0$ and $\delta f$. Thus, we next consider some examples.

\subsection{Uncorrelated perturbations} \label{sec:uncorr}

First, consider perturbations that are uncorrelated with the background distribution. In particular, consider a separable function
\begin{align}
\delta f = M(f_0, \partial f_0/\partial X, ...) R(X) \, ,
\end{align}
 where $M(f_0, \partial f_0/\partial X, ...)$ is a function of $f_0$ and its derivatives, and $R(X)$ is the ``random'' component satisfying $\langle R \rangle = 0$ but $\langle R^2 \rangle > 0$. Here, brackets indicate an average over phase space, $\langle g \rangle = (1/V_{\mathcal P}) \int dX \, g$. We assume that the random component is uncorrelated with the background distribution and its derivatives,
 \begin{align}
 \langle g R \rangle &= 0 \, , \nonumber \\
 \langle g R^2 \rangle &= \langle g\rangle \langle R^2 \rangle \, ,
 \end{align}
for any function $g(f_0, \partial f_0/\partial X, ...) $. Then the first-order perturbation to $S_h$ vanishes, while the second-order term becomes:
\begin{align}
\frac{\Delta S_h}{S_h}  \sim - &\frac{1}{2} \langle R^2 \rangle  \frac{S_h[f_0]}{N^2} \frac{\int dX \, [h(f_0) f_0]'' M^2(f_0, \partial f_0/\partial X, ...) }{h'(N/S_h[f_0])}  \le 0 \, . \label{eq:sperturbrandom}
\end{align}
For scale-invariant weights $h = f^{q-1}$, this reduces to
\begin{align}
\frac{\Delta S_q}{S_q}  \sim - \frac{q}{2} \langle R^2 \rangle \frac{S_q^{q-1}[f_0]}{N^q} \int dX \, f_0^{q-2} M^2(f_0, \partial f_0/\partial X, ...) \, . \label{eq:sqperturbrandom}
\end{align}
Thus, fluctuations that are uncorrelated with the background distribution always decrease the dimensional entropies; intuitively, such fluctuations create additional substructure in the distribution. The relative decrease of entropy does not depend on the details of the random component of the perturbation (e.g., extent, shape), but only on its {mean square} amplitude $\langle R^2 \rangle$. The factor of $M$ will determine at which weight the relative change in entropy is maximized. For the simplest case of $M = f_0$, Eq.~\ref{eq:sqperturbrandom} reduces to
\begin{align}
\frac{\Delta S_q}{S_q}  \sim - \frac{q}{2} \langle R^2 \rangle \, .
\end{align}
Thus, in this case, since the fluctuations are strongest at the peak of the distribution, $\Delta S_q/S_q$ is minimized for $q \to \infty$. For other choices of $M$ (e.g., proportional to $\partial f_0/\partial X$), $\Delta S_q/S_q$ may be minimized at finite $q$.

\subsection{Perturbation of a uniform distribution}

Another example is a perturbation $\delta f$ to the uniform distribution $f_0 = f_c = N/V_{\mathcal P}$ {on a bounded domain}, which can be viewed as a special case of the example in Sec.~\ref{sec:uncorr} . Then the first-order term of the expansion in Eq.~\eqref{eq:taylor1} vanishes upon integration, leaving the second-order contribution:
\begin{align}
\frac{\Delta S_h}{S_h} \sim - \frac{1}{2} \frac{\left[ f_c h(f_c) \right]''}{h'(f_c)} \frac{1}{V_{\mathcal P}} \int dX \, \left( \frac{\delta f}{f_c} \right)^2 \le 0 \, . \label{eq:uniperturb}
\end{align}
In this case, {$\Delta S_h/S_h$ only depends on the mean square amplitude of the perturbation $\delta f$}; the shape or extent of the perturbation does not have an effect. This reduction is a consequence of the degeneracy of the uniform distribution, which does not have any nontrivial level sets to act as a reference scale. For the scale-invariant entropies, Eq.~\eqref{eq:uniperturb} simplifies to
\begin{align}
\frac{\Delta S_q}{S_q}  \sim - \frac{q}{2} \frac{1}{V_{\mathcal P}} \int dX \, \left( \frac{\delta f}{f_c} \right)^2 \, .
\end{align}
The function $\Delta S_q/S_q$ thus approaches a minimum as $q \to \infty$, indicating that higher weights (measuring large values of $f$) are inherently more sensitive to the fluctuation. 

\subsection{Perturbation of exponential distribution} \label{sec:expperturb}

From the previous two examples, it is evident that perturbations to an unstructured or uncorrelated distribution will decrease entropy due to the absence of the first-order contribution from the perturbation. To illustrate the effect of a {\it correlated} perturbation on a {\it structured} distribution, where the first-order term [Eq.~\eqref{eq:sperturb1}] dominates and enables nontrivial results, we now consider perturbations to an exponential distribution on ${\mathcal P} = \mathbb{R}_{\ge 0}$. For $x \in \mathbb{R}_{\ge 0}$, let $f(x) = f_0(x) + \delta f(x)$ with
\begin{align}
f_0 &= \frac{N}{x_0} e^{-x/x_0} \, , \label{eq:f0exp} \\
\delta f &= \epsilon \left[ - \cos{(k x)} + \frac{1}{1+ k^2 x_0^2} \right] f_0 \, ,
\end{align} 
where $x_0 > 0$ is the width of the unperturbed distribution, $\epsilon$ is the small perturbation amplitude ($|\epsilon| \ll 1$), and $k > 0$ is the wavenumber of the perturbation. The prescribed perturbation $\delta f$ satisfies $\int dx \, \delta f = 0$. For $\epsilon > 0$, $\delta f$ is negative near the peak ($x \ll \pi/k$), positive at $x \sim \pi/k$, and exponentially small at $x \gg \pi/k$ as long as $k x_0 \lesssim 1$. Thus, $\delta f$ broadens the distribution at a characteristic width of $x \sim \pi/k$. For $\epsilon < 0$, $\delta f$ instead narrows the distribution at a similar width.

The general form of $S_h$ is nontrivial to reduce analytically, but the scale-invariant dimensional entropies $S_q$ is tractable. Since the phase space is one-dimensional for this problem, the dimensional entropies represent characteristic scales in $x$; an explicit calculation using Eq.~\ref{eq:f0exp} yields $S_q[f_0] = x_0 q^{1/(q-1)}$. Using Eq.~\eqref{eq:plperturb1}, the perturbed entropy can be calculated to first-order as
\begin{align}
\frac{\Delta S_q}{S_q} &\sim \frac{\epsilon k^2 x_0^2 (1 + q) q}{(1 + k^2 x_0^2 ) (q^2 + k^2 x_0^2 ) } \, .
\end{align}
For $\epsilon > 0$, the entropy increases for all $q$ (as a consequence of the perturbation broadening the distribution). Conversely, for $\epsilon < 0$, the entropy decreases for all $q$.

The weight $q$ at which the relative change in entropy is extremized, which we denote $q_{\rm m}$, can be found by $(d/dq)  (\Delta S_q/S_q) |_{q_{\rm m}} = 0$, yielding
\begin{align}
q_{\rm m} &= k x_0 \left[ k x_0 + (1 + k^2 x_0^2)^{1/2} \right] \, .
\end{align}
Thus, $q_{\rm m}$ is a monotonically increasing function of $k$, confirming that broader perturbations mainly impact entropies at lower weights (smaller $q$). In particular, $q_{\rm m} \sim k x_0$ when $k x_0 \ll 1$.

The value of $S_q$ at the extremal weight $q = q_{\rm m}$ is 
\begin{align}
\frac{S_{q_{\rm m}}[f_0]}{x_0} = q_{\rm m}^{1/(q_{\rm m}-1)} = \left( k x_0 [ k x_0 + (1 + k^2 x_0^2)^{1/2} ] \right)^{1/\left[k x_0 [ k x_0 + (1 + k^2 x_0^2)^{1/2} ] - 1 \right]} \, . \label{eq:sqmax}
\end{align}
For $k x_0 \ll 1$, this reduces to $S_{q_{\rm m}} \sim 1/k$. This is the primary result of this work: {\it the dimensional entropy $S_q$ which is most strongly affected by the perturbation is the one that has a similar phase-space scale as the perturbation}.

In the opposite limit of narrow fluctuations, $k x_0 \gg 1$, Eq.~\eqref{eq:sqmax} indicates $S_{q_{\rm m}} \sim x_0 + \log{(2 k^2 x_0^2)}/(2 k^2 x_0)$. In this case, it is reasonable to subtract off the asymptotic value to recover a scaling $S_{q_{\rm m}} - S_{\infty} \sim \log{(2 k^2 x_0^2)}/(2 k^2 x_0)$ in the limit $k x_0 \gg 1$. Thus, the correspondence between the dimensional entropy and phase-space perturbation scale is not linear in this limit.

While the example described in this section is idealized by choosing $f_0$ and $\delta f$ that can be conveniently calculated by analytical methods, it seems reasonable to conjecture that for a broad class of structured distributions with imposed perturbations, $\Delta S_q/S_q$ will be maximized for $S_q$ that lie at similar phase-space scales as the perturbation. This provides a connection between dynamics (which perturb $f$) and generalized entropies, which may be exploited in phenomenological models.

\section{Applications} \label{sec:apps}

\subsection{Diagnostic for irreversibility (in conservative systems)} \label{sec:vlasov}

The dimensional entropies described in this paper may be useful as a diagnostic for irreversibility, under the presumption that systems will tend to relax to high entropy (smooth, featureless) states. Conservative systems, in which the microscopic dynamics preserve the phase-space occupation volume of the level sets associated with the distribution function, form an example where generalized entropies may be relevant. To illustrate this, consider Boltzmann-type equations for the evolution of a continuous particle distribution $f(X;t)$ in the phase space ${\mathcal P} = \mathbb{R}^d$, over times $t \in \mathbb{R}_{\ge 0}$:
\begin{align}
\frac{\partial f}{\partial t} + {\mathcal F} \cdot \frac{\partial f}{\partial X} = {\mathcal C}[f] \, , \label{eq:vlas}
\end{align}
where ${\mathcal F}(X) \in \mathbb{R}^d$ is a phase-space flux (representing a ``force'' or ``flow velocity'') that is incompressible,
\begin{align}
\frac{\partial}{\partial X} \cdot {\mathcal F} = 0 \, ,
\end{align}
 and ${\mathcal C}[f](X)$ is a collision operator. For ${\mathcal C} = 0$, Eq.~\eqref{eq:vlas} becomes the Vlasov equation, which for appropriate boundary conditions (e.g., periodic or decaying to zero) ideally conserves all of the generalized entropies defined by Eq.~\eqref{eq:sdef} at the fine-grained level. In particular, $dS_h/dt = 0$ for ${\mathcal C} = 0$ follows from 
\begin{align}
\frac{d}{dt} \int dX \, f h(f) &= \int dX [f h(f)]' \frac{\partial f}{\partial t} \nonumber \\
&= - \int dX \, [f h(f)]' \left( {\mathcal F} \cdot \frac{\partial f}{\partial X} - {\mathcal C}[f] \right) \nonumber \\
&= - \int dX \, \left( {\mathcal F} \cdot \frac{\partial}{\partial X} [f h(f)] - [f h(f)]' {\mathcal C}[f] \right) \nonumber \\
&= - \int dX \, \frac{\partial}{\partial X} \cdot \left[ {\mathcal F} f h(f) \right] + \int dX \, [f h(f)]' {\mathcal C}[f] \nonumber \\
&=  \int dX \, [f h(f)]' {\mathcal C}[f]  \, , \label{eq:cas}
\end{align}
where the integral with the ${\mathcal F}$ term becomes zero by the divergence theorem with the appropriate boundary conditions. When $f$ is coarse-grained across finite scales in $\mathcal{P}$, however, the generalized entropies may not be conserved. Intuitively, microphysical degrees of freedom allow the macroscopic evolution to break generalized entropy conservation.

As an example, consider the case of entropy in a collisionless (or weakly collisional) plasma, as measured {\it in-situ} by spacecraft (in the solar wind or Earth's magnetosphere) or in particle-in-cell simulations \citep{liang_etal_2019, liang_etal_2020, pezzi_etal_2021, jara-almonte_ji_2021, cassak_etal_2023}. Our previous work introduced dimensional entropies for the (relativistic) Vlasov-Maxwell system of equations used to described a collisionless plasma \citep{zhdankin_2022, zhdankin_2022b}. In this case, ${\mathcal P} = \mathbb{R}^6$ for a population of $N$ particles with three-dimensional coordinates in both space and momentum, such that $X = (\boldsymbol{x},\boldsymbol{p})$. The phase-space flux becomes ${\mathcal F} = (\boldsymbol{v}, \boldsymbol{F})$ where $\boldsymbol{v} = \boldsymbol{p} c/(m^2 c^2 + p^2)^{1/2}$ is the (relativistic) particle velocity for the plasma species having rest mass $m$, and $\boldsymbol{F}(\boldsymbol{x},\boldsymbol{p})$ is the Lorentz force containing electromagnetic terms. In this scenario, momentum space is infinite but constrained by finite average energy $E = \int d^3x d^3p \, (m^2 c^2 + p^2)^{1/2} f(\boldsymbol{x},\boldsymbol{p})$, while homogeneity and isotropy imply no preferred spatial location $\boldsymbol{x}$ or direction of $\boldsymbol{p}$ (although note that the presence of a background magnetic field would break isotropy). In this case, only nontrivial dimension of the phase-space volume is the magnitude of the momentum, $p = |\boldsymbol{p}|$. As described in Ref.~\citep{zhdankin_2022b}, the scale-invariant dimensional entropies can then be cast in the form of characteristic momenta, which were called the ``Casimir momenta'', taking the form
\begin{align}
p_q[f] = n_0^{1/3} \left( \frac{1}{N} \int d^3x d^3p \, f^q \right)^{-1/3(q-1)} \, ,
\end{align}
where $n_0 = N/V_x$ is the number density of particles in the physical space of volume $V_x$. The physical significance of $p_q$ is that it describes the typical extent of the distribution in momentum space, and the evolution of $p_q$ indicates entropy production at those energies. While the Vlasov equation predicts $p_q[f]$ to be conserved at the fine-grained level, the corresponding quantity $p_q[\overline{f}]$ for the coarse-grained distribution function $\overline{f}(\boldsymbol{x},\boldsymbol{p})$ will not necessarily be conserved, at any coarse-graining scale \citep{eyink_2018}. Additionally, phenomena such as shocks or entropy cascades may lead to collisional anomalies \citep{schekochihin_etal_2009, tatsuno_etal_2009, adkins_etal_2018, kawamori_yu-ting_2022}. Particle-in-cell simulations of relativistic turbulence confirm that $p_q$ grow at a rate comparable to the average momentum as energy is injected into the system \citep{zhdankin_2022b}, which informs models of nonthermal particle acceleration \citep{zhdankin_2022, ewart_etal_2023}. Local regions of high entropy production rate also reveal the locations of dissipative structures. This application encourages further use of dimensional entropies for diagnosing irreversibility in conservative systems. Generalized entropies may also be useful for constructing fluid models of collisionless plasmas \citep[e.g.,][]{most_etal_2022}.

\subsection{Generalized maximum entropy modeling}

The principle of maximum entropy plays a central role in deriving equilibrium distributions in statistical mechanics, with strong connections to information theory \citep{jaynes_1957}. Generalized entropies were previously considered as a foundation for nonequilibrium statistical mechanics \citep[e.g.,][]{tsallis_1988, plastino_plastino_1999, hanel_etal_2014, jizba_korbel_2019}, although the procedure has a number of formal and conceptual issues. In this subsection, we derive the maximum-entropy distribution associated with $S_h$ subject to constraints. In principle, there may exist an even broader class of maximum-entropy distributions derived from the composite entropies discussed in Sec.~\ref{sec:composite}, but these are nontrivial to study with analytical methods.

The maximization of $S_h$ for a given function $h$ yields the same result as extremizing the more general kernel-form entropies $S_{H,h}[f]$ from Eq.~\ref{eq:sgen} (with appropriate conditions on $H$, such as monotonicity) \citep{jizba_korbel_2019}. In other words, for the purposes of deriving maximum-entropy distributions, the form of the regularization function $H$ is irrelevant. Here we choose to maximize $S_{h}[f]$ at fixed $N$ and other constraints $\Phi_i = \int dX \, f \phi_i/N$, $i \in \{1, \dots, N_\phi\}$, where $\phi_i(X) \ge 0$ are a set of $N_\phi$ constraint functions. Then we can write the action
\begin{align}
{\mathcal L} &= \frac{N}{h^{-1}\left( \frac{1}{N} \int dX \, f h(f) \right)} + \lambda_0 \left( N - \int dX \, f \right) + \sum_{i=1}^{N_\phi} \lambda_i \left( N \Phi_i- \int dX \, f \phi_i  \right) \, , \label{eq:lang}
\end{align}
where $\lambda_i$ (with $i \in \{ 0, \dots N_{\phi} \}$) are the Lagrange multipliers. Requiring $\delta {\mathcal L} / \delta f = 0$ for any $\delta f$, assuming $h f \to 0$ and $f \phi_i \to 0$ as $f \to 0$, we obtain the maximum-entropy distribution
\begin{align}
f_h(X) &= h^{-1}{\left( C_0 + \sum_{i=1}^{N_\phi}  C_i \phi_i(X) \right)} \, , \label{eq:fh}
\end{align}
where $C_0$ and $C_i$ are constants that can be determined by the normalization $\int dX \, f_h = N$ and constraints $\int dX \, f_h \phi_i = N \Phi_i$. Since $h$ was assumed to be invertible, $h^{-1}$ is monotonic.

In many physical systems, the primary constraint is total energy $N \langle E \rangle = \int dX \, f E$ where $E(X) \ge 0$ is the energy function and angle brackets indicate the average value. Consider such a system where $E(X)$ takes a minimum value of zero and may extend to infinity. Let $f_{\rm g}$ denote the value of $f_h$ at $E(X)=0$, being the ``ground state''; $f_{\rm g}$ will typically equal $f_{\rm max}$. Then $C_0 = h(f_{\rm g})$, and $C_1$ can be determined in terms of $S_h[f_h]$ by using Eq.~\ref{eq:sdef}. As a result, Eq.~\ref{eq:fh} can be expressed in the form
\begin{align}
f_h(X) &= h^{-1}{\left( h(f_{\rm g}) + \left[ h(N/S_h) - h(f_{\rm g})\right] \frac{E(X)}{\langle E \rangle}  \right)} \, . \label{eq:fhe}
\end{align}
For phase-space regions lying at the average energy, $E(X) = \langle E \rangle$, the distribution evaluates to $f_h(X) = N/S_h[f_h]$. Thus, for the maximum-entropy distribution $f_h$, the maximized entropy $S_h$ indicates the inverse phase density $N/f_h$ at the average energy $\langle E \rangle$. 

To be realizable physically, the maximum-entropy distribution must satisfy $f_h \ge 0$. If $h(f)$ diverges as $f \to 0$, then $f_h \ge 0$ for all energies. However, if $h(f)$ approaches a constant (or zero) as $f \to 0$, then there is a critical energy at which Eq.~\eqref{eq:fhe} implies $f_h = 0$, given by
\begin{align}
E_{\rm crit} \equiv \frac{h(0) - h(f_{\rm g})}{h(N/S_h) - h(f_{\rm g})} \langle E \rangle \, .
\end{align}
For $E(X) > E_{\rm crit}$, $f_h$ becomes ill-defined. In this situation, one must impose $f_h = 0$ for $E(X) > E_{\rm crit}$, with the implicit assumption in the maximization procedure being that the integration volume in Eq.~\eqref{eq:lang} is limited to phase-space regions where $f > 0$.

As an example, for the standard BGS entropy, taking $h(f) \propto \log{f}$, Eq.~\ref{eq:fhe} becomes
\begin{align}
f_{\log{f}}(X) = f_{\rm max} \exp{\left[ - \log{(f_{\rm max} S_{\log{f}}/N)}\frac{E(X)}{\langle E \rangle}  \right]} \, . \label{eq:flog}
\end{align}
It may be useful to compare Eq.~\eqref{eq:flog} to classical thermodynamics, for a thermodynamic system of $N$ non-relativistic particles in momentum space: ${\mathcal P} = \mathbb{R}^3$, $X = \boldsymbol{p}$, and $E(\boldsymbol{p}) = \boldsymbol{p}^2/2m$ where $m$ is the particle mass (assumed identical for all particles). The Maxwell-Boltzmann distribution takes the form
\begin{align}
f_{\rm MB}(\boldsymbol{p}) = \frac{N}{(2 \pi m T)^{3/2}} \exp{\left( - \frac{E(\boldsymbol{p})}{T}  \right)} \, ,
\end{align}
where $T$ is the thermodynamic temperature (we set the Boltzmann constant $k_B=1$). Therefore $f_{\rm max} = N/(2 \pi m T)^{3/2}$, $T = \langle E\rangle/\log{(f_{\rm max} S_{\log{f}}/N)}$. Since $\langle E \rangle = (3/2) T$, the maximized dimensional entropy can be expressed as $S_{\log{f}} = e^{3/2} N/f_{\rm max} = (2 \pi e m T)^{3/2} = (2 \pi e/3)^{3/2} p_{\rm rms}^3 \approx 13.6 p_{\rm rms}^3$ where $p_{\rm rms} = \langle p^2 \rangle^{1/2}$ is the rms momentum spread. Thus, $S_{\log{f}}$ has the interpretion of being the volume $(4\pi/3) p_c^3$ contained by a sphere of radius $p_c = e^{1/2} (\pi/6)^{1/6} p_{\rm rms} \approx 1.48 p_{\rm rms}$ in momentum space.

As another example, for the scale-invariant weight function, $h(f) = f^{q-1}$, Eq.~\eqref{eq:fhe} becomes
\begin{align}
f_q(X) &= f_{\rm max} \left( 1 + \left[ \left( \frac{N}{S_q f_{\rm max}} \right)^{q-1}  - 1  \right] \frac{E(X)}{\langle E \rangle}  \right)^{1/(q-1)} \, . \label{eq:fpl}
\end{align}
When $q > 1$, we must also set $f_q = 0$ when $E(X) > E_{\rm crit} = \langle E \rangle/[1-(N/S_q f_{\rm max})^{q-1}]$; in this case, $f_q$ is flat at low energies and has a steep drop to zero when $E \to E_{\rm crit}$. When $q < 1$, $f_q$ has a power-law tail at high energies, $f_q \propto E^{-\alpha}$, with the index $\alpha = 1/(q-1)$ varying from $1$ (when $q \to 0$) to $\infty$ (when $q \to 1$). When $q \to 1$, $f_q \to f_{\log{f}}$. Eq.~\eqref{eq:fpl} is equivalent to the Tsallis distribution \citep{tsallis_1988} and the kappa distribution known from the space and plasma physics communities \citep{pierrard_lazar_2010,livadiotis_mccomas_2013}.

\section{Conclusions} \label{sec:conclusions}

In this work, we introduced a dimensional framework for representing generalized entropies with generic weight functions, demonstrating that a broad class of functionals can be interpreted as entropies with physical dimensions of phase-space volume. These generalized entropy measures [Eqs.~\eqref{eq:sdef}, \eqref{eq:sqdef}, and \eqref{eq:mostgen}] satisfy the first three Shannon-Khinchin axioms, and describe the characteristic phase-space volume occupied by level sets of the distribution function. In particular, we constructed the {\it dimensional entropies} (Eq.~\ref{eq:sdef}):
\begin{align}
S_h[f]  \equiv \frac{N}{h^{-1}{\left( \frac{1}{N} \int dX \, f h(f) \right)}} \, ,
\end{align}
where $h$ must satisfy conditions on curvature [$f h(f)$ is convex if $h'(f)>0$ or concave if $h'(f)<0$] and convergence [$f h(f) |_{f=0} = 0$]. {As a special case}, we described the {\it scale-invariant dimensional entropies} (Eq.~\ref{eq:sqdef}):
\begin{align}
S_q[f] \equiv S_{f^{q-1}}[f] = N \left( \frac{1}{N} \int dX \, f^q \right)^{-1/(q-1)} \, ,
\end{align}
where $q > 0$. Notably, $S_q$ form a unique subset of $S_h$ that requires no internal dimensional parameters and thus is invariant to the normalization of $f$. Since $S_h$ are dimensional, they can be connected to processes occurring in the corresponding regions of phase space. In principle, the set of $\{ S_h \}$ can be used as a basis to construct composite entropies of increasing complexity (such as Eq.~\ref{eq:mostgen}). We described the mathematical properties, perturbative expansion, and corresponding maximum-entropy distributions of $S_h$. Critically, in Sec.~\ref{sec:expperturb}, we demonstrated that correlated perturbations on a structured distribution will have the largest effect on dimensional entropies that are at a similar phase-space scale, as illustrated for the case of an exponential distribution. This confirms the link between dynamics (perturbations) and statistics (entropies).

The set of dimensional entropies $\{ S_h \}$ contain many of the generalized entropies that exist in the literature \cite[such as ones reviewed by Ref.~][]{ilic_etal_2021}, in the sense that the weight function $h(f)$ may share the same form. The main novelty of the present formulation is that these functionals are manipulated into a dimensional form that can be interpreted easily, without requiring arbitrary normalization factors.

{In addition to applications for statistical physics (as described in Sec.~\ref{sec:apps}), the scale-invariant subsets of the dimensional entropies may be of mathematical interest for understanding fractals and multifractals (associated with, e.g., strange attractors of chaotic systems), which were previously tied to R\'{e}nyi and Tsallis entropies \citep[e.g.,][]{hentschel_procaccia_1983, alemany_zanette_1994, jizba_arimitsu_2004}. In particular, dimensional entropies provide an explicit link between the volume filling fraction of multifractals and scaling indices of generalized dimensions.}

Specific applications of the dimensional entropies will be considered in future work. Generalized entropies are traditionally challenging to apply due to the infinite number of possible weight functions and definitions. In nonequilibrium statistical mechanics, this may reflect the non-universality of the statistics due to a dependence on the particular irreversible processes acting on the system. However, generalized maximum-entropy distributions form only a small subset of all possible distributions, so they may be a versatile framework for reduced models (especially in high-dimensional systems, ${\mathcal P} = \mathbb{R}^d$ with $d \gg 1$). Importantly, if the mechanisms of entropy production are understood adequately from first-principles modeling, then the relevant subsets of generalized entropy may be inferred.

\begin{acknowledgments}
Research at the Flatiron Institute is supported by the Simons Foundation.
\end{acknowledgments}


%

\end{document}